\documentclass[twocolumn,showpacs,preprintnumbers,amsmath,amssymb]{revtex4}
\usepackage{graphicx}
\usepackage{dcolumn}
\usepackage{bm}
\usepackage[usenames]{color}
\def\lamno { \mathrm{LaMnO_3} }

\begin{document}

\title{Electronic structure of A-type antiferomagnetic  $\mathbf{LaMnO_3}$ by GW approximation}
\author{Y. Nohara$^{1}$}
\author{A. Yamasaki$^{2}$}
\author{S. Kobayashi$^{3}$}
\author{T. Fujiwara$^{1,4}$}

    \affiliation{$^{1}$ Department of Applied Physics, University of Tokyo, Tokyo 113-8656, Japan\\
                 $^{2}$ Max-Planck Institut f\"ur Festk\"orperforschung, Heisenbergstrasse 1, Stuttgart D-70569, Germany\\
                 $^{3}$ Texas Instruments Japan Ltd., 6-24-1 Nishi-Shinjuku, Shinjuku-ku, Tokyo 160-8366, Japan\\ 
                 $^{4}$ Core Research for Evolutional Science and Technology  (CREST-JST), Japan Science and Technology Agency}
   
\date{\today}
\begin{abstract}
The electronic structure of the A-type antiferromagnetic insulator 
$\lamno$ is investigated by the GW approximation. 
The band gap and spectrum are in a good agreement with experimental 
observation. 
The on-site/off-site dynamical screened Coulomb interaction on Mn sites is also calculated. 
Comparison with the results by the Hartree-Fock and the static COHSEX approximations 
is discussed in detail. 
The dynamical screening effect is important for the band gap and widths.  
\end{abstract}
\maketitle

Transition metal oxides show particular characteristics of physical properties, 
especially  large controllable change of properties with a small change of 
electron/hole doping or the external field. 
Among them, the hole-doped manganese oxides of perovskite structure,
A$_{1-x}$B$_{x}$MnO$_{3}$ (A$=$La$^{3+}$, Pr$^{3+}$, Nd$^{3+}$, $\cdots$ and  B$=$Ca$^{2+}$, Sr$^{2+}$, Ba$^{2+}$,  $\cdots$),
have been paid much attention  because of their properties, 
such as colossal magneto resistance (CMR), 
orbital, spin and charge order, metal-insulator transition 
and others.~\cite{hole-doped_P}
The starting material $\lamno $ is A-type antiferromagnetic (A-AF) 
at low temperatures. 
The Mn$^{3+}$ ion has t$^3_{2{\rm g}}$e$^1_{\rm g}$ configuration,
whose local spin quantum number $S$=2 caused by the Hund coupling. 
The atomic spin couples ferromagnetically with those of neighboring Mn ions 
on a same basal plane and couples antiferromagnetically with those 
of Mn ions on different planes.~\cite{wk-1955} 
The system shows the C-type orbital ordering with the Jahn-Teller (JT) 
distortion.~\cite{A-Type_Mn} 
The local spin density approximation (LSDA) is powerful tool 
to investigate the electronic structures. 
However, the LSDA sometimes does not give an overall picture of the electronic structures 
in transition metal compounds with strong correlation.  
In fact, the electronic structures of $\lamno $ were calculated by 
the LSDA and  the LSDA+U methods 
and these methods reveal several problems. 
For example, the LSDA underestimates the band gap 
and overestimates d-band width.~\cite{LSDA_LaMnO3} 
In the LSDA+U method,~\cite{LDA+U_LaMnO3-1,LDA+U_LaMnO3-2}  
the Hubbard-type Coulomb interaction $U$ is introduced so that 
the resultant band gap agrees with the observed optical gap. 
However, this method produces a strong bonding state 
with O-p and Mn-d orbitals 
in the deep valence region as an artifact.  
The GW approximation (GWA) is a promising method 
for systems with electron-electron correlation 
but requires much heavier computational task. 
The GWA was applied to $\lamno$ with a smaller unit cell
in fictitious cubic paramagnetic phase,   
neglecting the JT distortion.~\cite{GWA_LaMnO3} 
However, the  JT distortion and 
tilted MnO$_6$ octahedra (GdFeO$_3$-type distortion) 
are crucially important in A-AF $\lamno$.~\cite{SHT-96} 
The electronic structure of A-AF $\lamno$  has not been calculated so far,  
because the unit cell contains 20 atoms, {\it i.e.} 4 $\lamno$,
and large memory space and the parallel computation are needed. 
In this letter, the electronic structure of 
the A-AF  $\lamno $ is investigated by the GWA, 
adopting the experimentally determined crystal structure 
of the orthorhombic $Pbnm$ symmetry with 
the JT and the GdFeO$_3$-type distortion.~\cite{Structure} 
Present calculation is 
based on the Linear Muffin-Tin Orbital method with 
the Atomic Sphere Approximation (LMTO-ASA).~\cite{LMTO}
We use the $4 \times 4 \times 4$ ${\bf k}$-mesh  
in the Brillouin zone  and the resultant 
30 ${\bf k}$-points in the irreducible zone.
The set of the maximum orbital angular momentum of the LMTO basis 
in La, Mn, O and empty spheres are chosen to be (fddp).
All $E_\nu$'s for the LMTO linearization energy points are fixed at the averaged 
energies of occupied bands for respective orbital except La-p, 
for which $E_\nu$ is chosen in the unoccupied region. 
The GWA is based on the many-body perturbation theory 
and is actually the first term approximation of the perturbation series  
for the one-body Green's function,~\cite{Hedin} 
where the self-energy is replaced by the lowest order term of 
the expansion as $\Sigma = i GW$. 
Here $G$ and $W$ are the one-particle Green's function and
the dynamically screened interaction, respectively. 
The dynamically screened interaction $W$ in the GWA is treated by 
the random phase approximation (RPA) as 
\begin{align}
 W = v + v\chi^0 W, \label{dyn-int2}
\end{align}
where 
$v$
is the bare Coulomb interaction and $\chi^0$ is the irreducible
polarization function $\chi^0 = -i GG$.

We presume the wave-functions $\{ \psi_{{\bf k}n}({\bf r})\}$ 
of the LSDA to be reasonably good starting wave-functions, and 
then we adopt the LSDA Hamiltonian $H_0=T+V^{\rm H}+V^{\rm xc}_{\rm LSDA}$ 
as an unperturbed one. 
Here $T$ is the kinetic energy, $V^{\rm H}$ is the Hartree potential,
and $V^{\rm xc}_{\rm LSDA}$ is the exchange-correlation potential in the LSDA.
The self-energy correction can be written with three terms as
$\Delta\Sigma = \Sigma^{\rm x} + \Sigma^{\rm c} - V^{\rm xc}_{\rm LSDA},$
where $\Sigma^{\rm x} (= i G v)$ is the exchange part (the Fock term) and 
$\Sigma^{\rm c} (= i G W^{\rm c})$ is the dynamical correlation part. 
$W^{\rm c}$ is the second term in eq.(\ref{dyn-int2}).
The Green's function is defined as 
\begin{equation}
G=(E-H_0-\Delta \Sigma)^{-1}
\end{equation}
and the resultant quasiparticle energy is given as
\begin{eqnarray}
 E_{{\bf k}n} 
 &&= \epsilon_{{\bf k}n} + Z_{{\bf k}n} 
  {\rm Re}\Delta\Sigma_{{\bf k}n} (\epsilon_{{\bf k}n}) ,
\label{GW-band}
\end{eqnarray}
where $\epsilon_{{\bf k}n}$ is the LSDA eigen-energy. 
The self-energy is 
$\Delta\Sigma_{{\bf k}n} (\epsilon_{{\bf k}n}) = 
 \langle \psi_{{\bf k}n} | \Sigma^{\rm x} + \Sigma^{\rm c}(\epsilon_{{\bf k}n}) -
 V^{\rm xc}_{\rm LSDA} | \psi_{{\bf k}n} \rangle$
and the renormalization factor is 
$Z_{{\bf k}n}= ( 1 - \partial {\rm Re}\Delta\Sigma_{{\bf k}n}(\epsilon_{{\bf k}n})
/ \partial E )^{-1}$. 
The renormalization factor $Z_{{\bf k}n}$ is an inverse effective mass ratio.
In the present work we perform one-iteration calculation without
self-consistency, {\it i.e.} 
$\Sigma=iG_0W_0$ and $\chi^0=-iG_0G_0$.

\begin{figure}[htbp] 
\begin{center}
\resizebox{0.46\textwidth}{!}{
\includegraphics[height=18cm,clip]{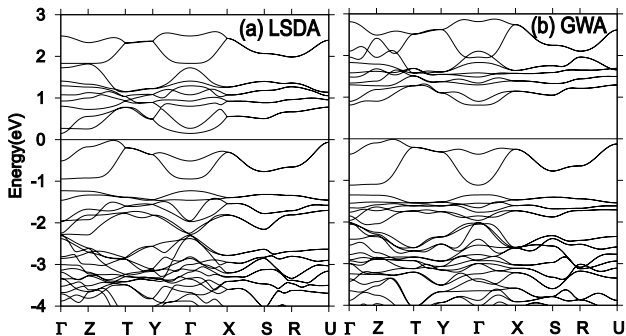}
}
\end{center}
\vspace{-5mm}
\caption{\label{fig:band.lda.gw} 
Energy band of the A-AF $\lamno$  by the LSDA (a), and by the GWA (b).
The energy zero-th is fixed at the top of the respective valence band.    
}
\end{figure}

The important steps of our GWA are the offset method,~\cite{KS-1}
the product-basis set~\cite{Ferdi-1} and parallel computation. 
The offset method is essential to reproduce a response function with 
a small momentum transfer $\hbar{\bf k}$ 
and we use nearby 6 ${\bf k}$ offset points  
$\frac{2\pi}{a}(\pm 0.0784\times \frac{a}{b}, 0, 0)$,
$\frac{2\pi}{b}(0, \pm 0.0784, 0)$, 
$\frac{2\pi}{c}(0, 0, \pm 0.0784\times \frac{c}{b})$ 
instead of ${\bf k}=0$, where $a$, $b$ and $c$ are the lattice constants 
in the orthorhombic system.  
In the calculation of the self-energy, a product of two spherical wave-functions  
$\phi\phi$ is expanded in terms of the new spherical functions called a product basis, 
whose total number is 1240 for the present choice.
The product bases are specified by the total angular momentum of two wave-functions 
and we adopt the maximum total angular momentum of the product bases 
up to (fddp) for the calculation of $\Sigma_{\rm c}$  
as the LMTO's themselves. 
This choice of total angular momentum reduces the number of the product bases  to 700. 
See also refs.\cite{AF-1,AF-2}. 

%
The band structures calculated by (a) the LDA  ($\epsilon_{{\bf k}n}$) and 
(b) the GWA ($E_{{\bf k}n}$) 
are shown in Figure~\ref{fig:band.lda.gw}. 
The band gap by the LSDA is as small as 0.16~eV and,  
on the other hand, that by the GWA becomes 0.82~eV, 
which agrees reasonably well with experimental optical gap 1.1~eV.~\cite{Optical_Gap} 
We can see that the width of the majority Mn-d$({\rm t_{2g}})$
band (in the energy region $-2.3~{\rm eV} \le \epsilon_{{\bf k}n} \le -1~{\rm eV}$ in 
the LSDA results) 
becomes narrower, and that of the majority Mn-d$_{3z^2-r^2}$ band 
(in the energy region $-1.0~{\rm eV} \le \epsilon_{{\bf k}n} \le 0~{\rm eV}$ 
in the LSDA results) becomes broader.
The La-d and La-f bands appear in the LSDA calculation at the energy range 
of $3~{\rm eV} \le \epsilon_{{\bf k}n} \le 7.5~{\rm eV}$ above 
the top of the valence band. 
In the GWA, these bands, La-d and La-f, are very strongly affected and  
shift to the energy range of $10~{\rm eV} \le \epsilon_{{\bf k}n} \le 15~{\rm eV}$. 
For the La-f, 
the band positions in the LSDA seem too low and also 
the LSDA Hamiltonian might not be a good starting point. 
%
%

\begin{figure}[thbp] 
\begin{center}
\resizebox{0.48\textwidth}{!}{
  \includegraphics[height=18cm,clip]{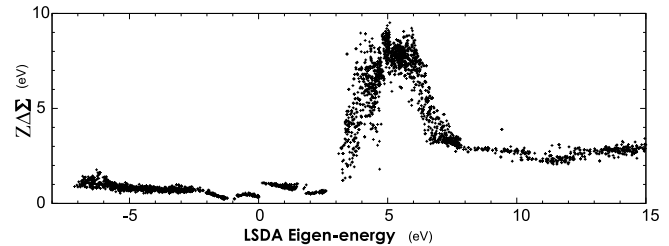}
}
\end{center}
\vspace{-5mm}
\caption{\label{fig:slf} 
The renormalized self-energy 
$Z_{{\bf k}n}{\rm Re}\Delta \Sigma_{{\bf k}n}(\epsilon_{{\bf k}n})$ 
as a function of the LSDA eigen-energies $\epsilon_{{\bf k}n}$. 
The energy zero-th 
is fixed at the top of the LSDA valence band. 
}
\end{figure}

%
The  renormalized self-energy $Z{\rm Re}\Delta \Sigma$ 
is shown in Figure~\ref{fig:slf} as a function of 
the LSDA eigen-energy $\epsilon_{{\bf k}n}$. 
A step can be seen at $\epsilon_{{\bf k}n}=0$  
and it makes the band gap wider.  
In an energy region where $Z{\rm Re}\Delta \Sigma$ is decreasing/increasing as a function of 
$\epsilon_{{\bf k}n}$, the width of these bands becomes 
narrower/broader.    
Thus, we find that, in the GWA, 
the O-p band in the energy range 
$-7.5~{\rm eV} \le \epsilon_{{\bf k}n} \le -2.5~{\rm eV}$ 
and the Mn-d(t$_{2g}$) band in the energy range 
$-2.5~{\rm eV} \le \epsilon_{{\bf k}n} \le -1.0~{\rm eV}$ 
become narrower.
The reduction of the O-p  band is attributed to 
that of hybridized Mn-t$_{2{\rm g}}$ component.  
The value of $Z$ is around 0.6 in the Mn-d band region 
($-2.5~{\rm eV}<E<3.0~{\rm eV}$) 
and around 0.7 in the region of the O-p valence bands.
%
%

\begin{figure}[tbhp] 
\begin{center}
\resizebox{0.46\textwidth}{!}{
\includegraphics[height=18cm,clip]{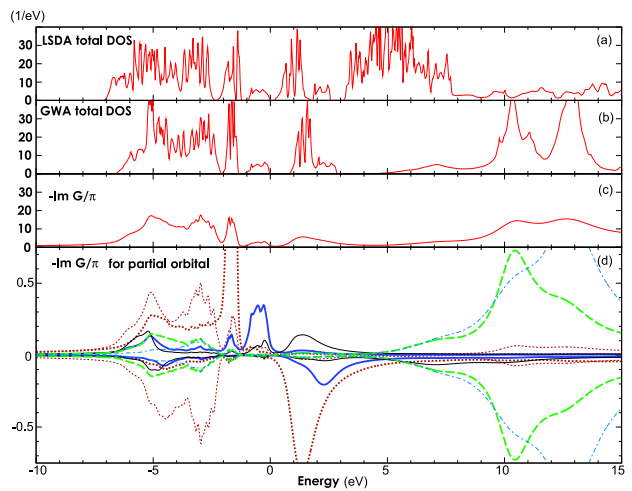}
}
\end{center}
\vspace{-5mm}
\caption{\label{fig:dos} 
(Color online) (a) Total DOS of the LSDA, (b) total DOS of the GW quasiparticles,  
and the imaginary part of the Green's function $-(1/\pi ){\rm Im}G$ for 
(c) the total and 
(d) the partial components per one respective atom. 
In (d), the upper and the lower panels show the partial components 
with majority and minority spins,  and
bold solid line: Mn-d$_{3z^2-r^2}$, 
thin solid line: Mn-d$_{x^2-y^2}$,
bold dotted line: Mn-d(t$_{2g})$, 
thin dotted line: O-p, 
bold chain line: La-d, 
dot-chain line: La-f. 
The enrgy zero-th in (a) is fixed at the top of the valence bands 
of the LSDA and  in (b)$\sim$(d) at that of the GWA.
}
\end{figure}
%

%
Figure~\ref{fig:dos} shows the density of states (DOS) or  
the imaginary part of the Green's function by the LSDA and 
the GWA. 
The band gap in the total DOS for the quasiparticle energy (b) is 0.82~eV. 
The imaginary part of the Green's function shows the smoothed and widely spread 
spectrum,  because a quasiparticle has a finite lifetime.
It should be noticed that the imaginary part of the Green's function 
gives an excellent agreement, except the La-d, with the observed 
XPS and XAS.~\cite{Spectra}
In Figure~\ref{fig:dos}(d), the notation of orbitals refers to the local coordinate.  
For example, d$_{3z^2-r^2}$ is the orbital along the longest Mn-O pair axis and 
d$_{x^2-y^2}$ on the perpendicular plane nearly along other Mn-O pair axes.  
The orbitals of d(t$_{2g}$) and d(e$_g$) are split by the cubic crystal field and 
those of d$_{3z^2-r^2}$ and d$_{x^2-y^2}$ mainly 
by the JT distortion.
The width for the valence bands does not change much and, on the other hand, 
that of the conduction bands becomes much broader.  
The partial Green's function is identical for all Mn ions, 
once one uses local orbitals in accordance with the local coordinates. 
This facts clearly show the characteristic orbital order 
(orbital C-type `antiferro'-coupling) in $\lamno$;  
{\it i.e.} each local orbital d$_{3z^2-r^2}$ aligns along the elongated 
Mn-O axis with antiferro- and ferro-coupling 
within and between basal planes, respectively. 
%
%

\begin{figure}[hbtp] 
\begin{center}
\resizebox{0.43\textwidth}{!}{
  \includegraphics[height=13cm,clip]{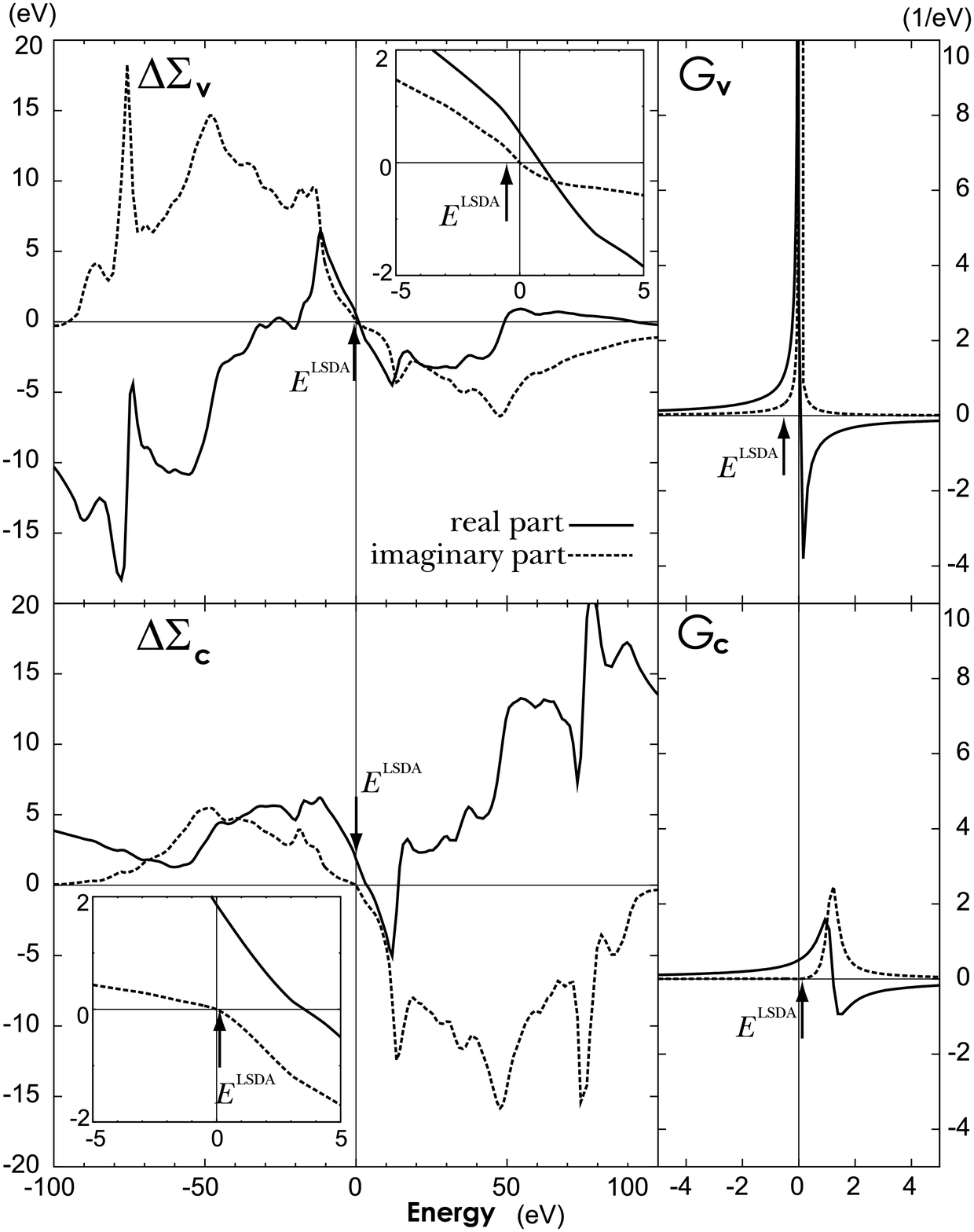}
}
\end{center}
\vspace{-5mm}
\caption{\label{fig:Gk} 
The self-energy (left) and Green's function (right) 
of the highest valence (top) and lowest conduction (bottom) 
bands at  ${\bf k}=0$
(solid line: real part, dotted line: imaginary part).
The energy zero-th is set at the top of the LSDA valence band. 
The inset shows the self-energy in a narrower energy range 
and the arrows indicate the position of the LSDA eigen-energies.
}
\end{figure}

Figure \ref{fig:Gk} shows the self-energies and the Green's functions 
of the top of the valence band (d$_{3z^2-r^2}$) and 
the bottom of the conduction band (majority d$_{x^2-y^2}$ $+$ minority d(t$_{2g})$) 
at $\Gamma$-point (${\bf k}=0$). 
Actually the overall feature of the self-energy is not very sensitive 
on the ${\bf k}$-points. 
One can notice the large difference of the shape of the Green's function;  
the one of the conduction band is much broader than that of the 
valence band. 
This fact is also confirmed for all deeper valence bands. 
The broadness of the Green's functions of the conduction bands 
is consistent to the behavior of the imaginary part of the self-energy.  
This fact is presumably attributed to the fact that the wave-functions 
of the conduction bands extent appreciably over the neighboring oxygen 
sites and are affected by strong polarization fluctuation. 

\begin{figure}[hbtp] 
\begin{center}
\resizebox{0.44\textwidth}{!}{
  \includegraphics[height=18cm,clip]{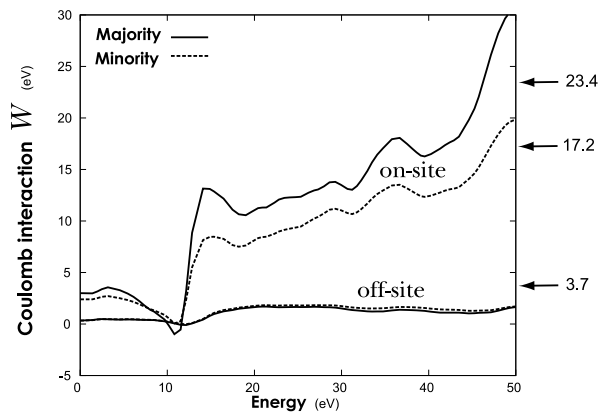}
}
\end{center}
\vspace{-5mm}
\caption{\label{fig:wlo} 
The dynamical screened Coulomb interaction in $\lamno$
(solid line: interaction with majority spins, 
 dotted line: interaction with manority spins). 
The arrows indicate the extremum values at the higher energies.}
\end{figure}

%
The magnitude of the on-site/off-site d-d Coulomb interaction 
on Mn sites 
$\langle \phi_{\rm d} \phi_{\rm d}|W(E)|\phi_{\rm d} \phi_{\rm d} \rangle$ 
is shown in Figure~\ref{fig:wlo}. 
The off-site one is for a nearest-neighbor Mn pair on a basal plane 
(ferromagnetically coupling pair) and 
the other one for a nearest-neighbor Mn pair on different planes 
(antiferromagnetically coupling pair) is almost identical.  
The Coulomb interaction is strongly screened 
at low energies and is reduced to a few eV; 
3.00~eV for on-site one with majority spins, 
2.40~eV for on-site one with minority spins, 
0.34~eV for off-site ones 
both with majority and minority spins. 
On the other hand, it approaches at the high energies  
to the values of bare ones 23.4~eV, 17.2~eV, 3.7~eV, respectively,   
because of no screening mechanism at high energy region. 

The screening effects depend upon the excitation energy, 
since the electrons participating in the screening mechanism 
depend very much on their excitation energy. 
The screening for the on-site Coulomb interaction is quite efficient 
and this phenomena is attributed to 
the screening by easily mobile e$_g$ electrons.~\cite{LDA+U_LaMnO3-2}  
The d(t$_{2g}$) wave-functions is more localized than the LSDA picture 
and the width of the Mn-d(t$_{2g}$) bands becomes narrower as seen in 
Figure~\ref{fig:band.lda.gw}(b).  
The Coulomb interaction with minority spins is smaller 
than that with majority spins, 
because the radial wave-function $\phi_{\rm d}(r)$ 
of minority spin has much components 
away from the core region 
and this fact results in a smaller values of a bare on-site Coulomb integral. 
It should be noticed that the screened Coulomb interaction is reduced to $W \simeq 0$ 
around $E \simeq 10$ eV  in Figure \ref{fig:wlo}.
The excitation from O-p occupied states to  La levels 
($3\sim 7.5$~eV)
hybridized with Mn-d causes strong screening effects  for Mn-d electrons. 
However, this large drop of the Coulomb interaction may be problematic, 
because the position of La-f bands may be 
too low in the LSDA calculation. 
Presumably, the large reduction of $W$ in the energy region of La-f or La-d bands 
could be true but it should  happen at higher energy region. 

\begin{table}[tbhp]
\caption{\label{Gap_moment_W} The band gap $E_{\rm G}$, the band widths 
of occupied d$_{3z^2-r^2}$, d($t_{2g}$) and O-p ($W_{3z^2-r^2}$, $W_{{\rm t}_{2g}}$
and $W_{\rm O_P}$) 
and the magnetic moment $M (\mu_\mathrm B)$. 
The band gap and widths are obtained from the calculated 
quasiparticle energy. 
The magnetic moments by the HFA, the COHSEX and the GWA 
are all the same. 
}
\begin{ruledtabular}
\begin{tabular}[t]{lccccc}
                         & LDA   & HFA   & COHSEX  & GWA    & expt.   \\ \hline
$E_{\rm G}$~(eV)         & 0.16  &10.04 &  0.71   & 0.82   &  1.1$^{a}$    \\
$W_{3z^2-r^2}$~(eV)      & 0.9   & 2.8  &  1.3    & 1.1    &   -     \\
$W_{{\rm t}_{2g}  }$~(eV)& 1.1   & 1.3  &  1.0    & 0.7    &   -     \\
$W_{\rm Op}$~(eV)        & 4.8   & 5.4  &  5.4    & 4.6    &   -     \\
$M (\mu_\mathrm B)$      & 3.49  &      &         &        &  3.7$^{b}$, 3.87$^{c}$    \\
\end{tabular}
\end{ruledtabular}
\footnotetext[1]{Ref.~\cite{Optical_Gap}.}
\footnotetext[2]{J.B.A.A. Elemans, B. van Laar, K.R. van der Veen and B.O. Loopstra, 
J. Solid State Chem. {\bf 3}, 238 (1971).}
\footnotetext[3]{Ref.~\cite{Structure}.}
\label{tab:results}
\end{table}%

%
%
We summarize in Table 1 the calculated results for band gap, 
widths of occupied bands and magnetic moment with experimental results for comparison.
The calculated results are for the LSDA, the Hartree-Fock approximation (HFA), 
the static COHSEX approximation (sCSA)~\cite{Hedin} and the GWA. 
The sCSA is obtained with the self-energy of $E=0$ 
or the zero excitation energy .
The band gap is found to be 0.82~eV by the GWA, 
which is in good agreement with experimental value 1.1 eV  
and slightly larger than the value by sCSA.
The large band gap of the HFA is owing to the approximation free 
from the self-interaction. 
Then the sCSA  adds the effect of the static screening. 
In fact, the change from the HFA to the cCSA is very 
large for the unoccupied Mn-d band. 
The dynamical part of the electron-electron interaction is
attributed to the small difference between the band gaps of  
the sCSA and the GWA.  
The magnetic moment of magnetic insulators is not changed from the results of the LSDA  in 
the HFA, the sCSA and the GWA, 
because these methods are the perturbative ones and 
we do not renew  wave-functions. 
The band widths of Mn-d$_{3z^2-r^2}$ states by the GWA and the sCSA  
are slightly greater than that of the LSDA. 
The band width of Mn-d(t$_{2g}$) states by the GWA and the sCSA  
are less than that of the LSDA.
The band width of O-p bands by GWA is less than the LSDA value. 
These results of the band widths indicate the wave-functions 
of Mn-d(t$_{2g}$) and O-p states becomes more localized by the GWA 
and that of Mn-d(e$_{g}$) less.   
Therefore, the screening effect affects to d(t$_{2g}$) and d(e$_g$) states 
in an opposite way.
In summary,  we applied the GWA to the A-type antiferromagnetic $\lamno $. 
Band gap and magnetic moment are in good agreement with experimental results.
The band width of Mn majority d(t$_{2g}$) states become narrower 
due to the correlation effect and, on the contrary, 
that of Mn majority d$_{3z^2-r^2}$ states becomes broader. 
The spectrum shows an excellent agreement with experimental results.
We also present the energy depending Coulomb interaction, 
and find that on-site d-d Coulomb interaction
is strongly screened at the low energy region 
due to mobile e$_g$ electrons. 
%
%

Computation was partially carried out by use of 
the facilities at the Supercomputer Center, Institute for Solid State Physics, 
University of Tokyo, and the Institute of Molecular Science at Okazaki.


\end{document}